\documentclass{aa}
\usepackage{epsf}
\usepackage{graphicx}
\usepackage{txfonts}

\begin{document}
   \title{V4332 Sagittarii revisited
         \thanks{Based on observations made at 
          the South African Astronomical Observatory}}

   \author{R. Tylenda\inst{1,3} \and L. A. Crause\inst{2} 
           \and S. K. G\'orny\inst{1} \and M. R. Schmidt\inst{1}}

   \offprints{R. Tylenda \email{tylenda@ncac.torun.pl}}

   \institute{Department for Astrophysics, 
              N. Copernicus Astronomical Centre,
              Rabia\'nska 8, 87-100 Toru\'n, Poland
           \and South African Astronomical Observatory, PO Box 9,
		Observatory 7935, South Africa
           \and Centre for Astronomy, N. Copernicus University,
                Toru\'n, Poland}

   \date{Received   }

   \abstract{The eruption of V4332~Sgr discovered in February 1994 shows
striking similarities to 
that of V838~Mon started in January 2002. The nature of these
eruptions is, however, enigmatic and unclear.
We present new photometric and spectroscopic data on
V4332~Sgr obtained in April--May 2003 at the SAAO. The obtained spectrum
shows an unusual emission-line component superimposed on an early M-type
stellar spectrum. The emission-line spectrum is of very low excitation and is
dominated by lines from neutral elemets (NaI, FeI, CaI) and molecular bands 
(TiO, ScO, AlO). 
We also analyse all
the observational data, mainly photometric measurements, available for V4332~Sgr.
This allows us
to follow the evolution of the effective temperature, radius and luminosity of
the object since February~1994 till 2003. 
We show that the observed decline
of V4332~Sgr can be accounted for by a gravitational contraction of 
an inflated stellar envelope.
The combined optical and infrared photometry in 2003 shows that apart from 
the M-type stellar component there is a strong
infrared excess in the $KLM$ bands. 
This excess was absent
in the 2MASS measurements done in 1998 but was probably starting to appear in 
$K$ in 1999 when the object was observed in the DENIS survey.
We interpret the results in terms of a stellar merger scenario proposed
by Soker \& Tylenda. The infrared excess is likely to be due to 
a disc-like structure which is either of protostellar nature or has been
produced during the 1994 eruption and stores angular momentum from the merger
event.
   }

   \maketitle

\section{Introduction \label{int}}

V4332 Sagittarii was discovered towards the end of February~1994 as an
apparent nova.  Subsequent spectroscopic observations however indicated that
this was not a classical nova.  It lacked spectral features characteristic of
novae while its spectral type, K3--4~III--I on March~4, was too late for a
nova at maximum and with time, contrary to classical novae, it evolved
toward lower and lower effective temperatures.
A detailed description and
analysis of available observational data for V4332~Sgr in eruption can be
found in Martini et~al. (1999, hereafter \cite{martini}).

Similar evolution has been observed for V838~Mon which was discovered in
eruption in early January~2002.  This object underwent more complex
photometric and spectral evolution, but in March-April~2002 its behaviour
was quite similar to that observed for V4332~Sgr (e.g. Munari et~al.
\cite{munari}, Crause et~al. \cite{crause}).  Therefore V838~Mon and
V4332~Sgr (and presumably also M31~RV -- see Mould et~al. \cite{mould}) are
considered a mysterious new class of eruptive variables.  As discussed in
Soker \& Tylenda (\cite{soktyl}), thermonuclear events, i.e. nova-type
runaway or late He-shell flash, cannot account for the observed evolution of
these objects.  Instead these authors propose that the eruptions can be due
to stellar merger events (see also Retter \& Marom \cite{retmar}).

V4332~Sgr has been forgotten by observers since its decline in June 1994. 
As far as we know, the only observations obtained between 1994--2003 are
2MASS measurements in May~1998 and DENIS photometry in September~1999. 
Motivated by the interest in V838~Mon, we took low resolution optical
spectra of V4332~Sgr in April~2003, followed by photomertic measurements in
May~2003.  We were surprised to find that the object displayed an unusual
emission-line spectrum, completely different to that of typical
emission-line objects such as planetary nebulae, symbiotic stars or T~Tauri
stars.  The strongest emission feature appeared to belong to the Na~I~D
lines, but later we realised that many emission lines are due to
molecular bands of TiO, AlO and ScO.  Banerjee et~al. (2003, hereafter
\cite{baner}) observed V4332~Sgr in near-IR in April--June~2003.  Their
spectra also show strong emission features which they attribute to
AlO.  More recently, in September~2003, Banerjee \& Ashok (2004, hereafter
\cite{banash}) observed the object in the optical and obtained a spectrum very
similar to ours. Very recently, when the original version of this paper
was already submitted for publication,
Banerjee et~al (2004, hereafter \cite{bva})
presented infrared spectra obtained in September~2003 and April~2004.
They have detected a water-ice absorption band and a CO band in emission.
Thus V4332~Sgr is now an unusual emission-line object that
deserves detailed observations. In this paper we present our observations
and analyse the available data on the evolution of V4332~Sgr since its
eruption in 1994.

\section{Observations  \label{observ}}

The observations of V4332 Sgr were made at the South African Astronomical
Observatory (SAAO).  The spectra were obtained on 3 April 2003 with the 
grating spectrograph with SITe CCD mounted at the Cassegrain focus, f/18, of the
1.9-meter telescope.  The total duration of the exposure was 900 seconds
with the spectral coverage 3500--7400~$\AA$ and an average resolution of
1000 using grating No. 7 (300 lines/mm).  The selected slit width was
300$\mu$m giving $1\farcs8$ projected on sky, comparable to seeing
conditions during the observations. The FWHM of the night sky lines
was 5.6~$\AA$. Two standard stars, CD$-32\degr9927$
and EG 274, were observed during the night.  This allowed us to calibrate
the spectrum in flux units while the wavelength calibration was secured by
observing CuAr lamp spectra.

The standard reduction procedures were applied to the spectra with the
MIDAS\footnote{MIDAS is developed and maintained by the European Southern
Observatory} longslit package.  These included bias subtraction, flat-field
correction, atmospheric extinction correction, wavelength and flux
calibration.  

Cousins $BVR_cI_c$ photometry was obtained  with the 1.0-meter telescope and the
1k~$\times$~1k~STE4~CCD camera on 21~May~2003.  The frames were
bias-subtracted, overscan-trimmed and flat-fielded using the CCDPROC package
within IRAF. Target observations were calibrated against Landoldt standard stars.
The results are given in Table~\ref{phot_t}. 
Note that the errors given in the table are based only on 
the photon noise, they do not take into
account external sources such as the calibration. Thus the true errors are 
probably larger than those quoted in the table. 

The results in Table~\ref{phot_t}
can be compared to those obtained by \cite{banash} on 29~September~2003.
Apart from the $B$ magnitude, which is uncertain in both measurements and 
differs by 0.4, 
the other magnitudes are consistent within $0.03-0.11$.
Thus we can conclude that the object
did not evolve significantly during the 4 month period in 2003.

\begin{table}
\centering
\caption{Cousins photometry of V4332~Sgr obtained on 21 May 2003}
\label{phot_t}
\begin{tabular}{c c l l}
\hline
  UT  &  Band  &  Magnitude & Error\\
\hline
  03:50  &  $B$  &  19.6   & $\pm$0.2   \\
  03:15  &  $V$  &  17.63  & $\pm$0.03  \\
  03:27  &  $R_c$  &  16.34  & $\pm$0.00  \\
  03:39  &  $I_c$  &  15.09  & $\pm$0.00  \\
\hline
\end{tabular}
\end{table}

\section{Spectrum  \label{spectr}}

\begin{figure*}
\centering
  \includegraphics[width=17.0cm]{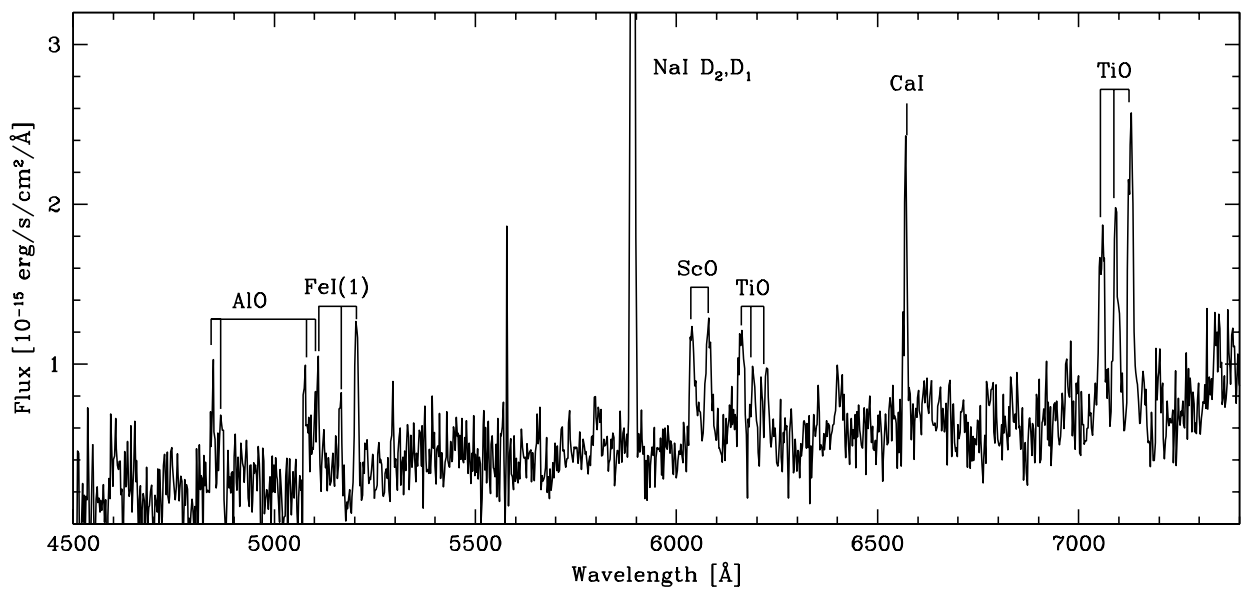}
  \caption{The spectrum of V4332~Sgr obtained on 3~April~2003.  The observed 
flux (in $10^{-15}$~erg~cm$^{-2}$~s$^{-1}$~$\AA^{-1}$) is shown versus the
wavelength (in \AA).  Due to the excessive observational noise, the spectrum 
below $\sim$4500\,\AA\ is not shown.  The sharp spike near 5580\,\AA\ is due 
to imperfect sky subtraction.  The labels identify the strongest emission 
features.}
  \label{spectr_f}
\end{figure*}

The resultant spectrum of V4332~Sgr obtained on 3~April~2003 is presented in
Fig.~\ref{spectr_f}.  The observational noise becomes excessive shortward of
 $\sim$4500\,\AA\ where the spectrum is faint so that section is not shown in 
Fig.~\ref{spectr_f}. For the remaining part a continuum level has been
fitted by the median filtering of the spectrum and the observational noise
has been derived by calculating the rms of the deviations from this
continuum, excluding the apparent emission features. The calculated average
noise is $0.19 \times 10^{-15}$~erg~cm$^{-2}$~s$^{-1}$~$\AA^{-1}$ 
and as a result the signal to
noise ratio (S/N) for the detected continuum changes from only 1 to $\sim$5 at
the longest wavelenghs. However for the emission lines the S/N can be
substantially higher. Our spectrum can be compared to that of
\cite{banash}.  Note, however, that our wavelength range does not extend far
enough into the red to register the prominent lines of KI.

As Fig.~\ref{spectr_f} shows, the spectrum is dominated by prominent emission
features.  This emission-line spectrum is, however, unusual and does not
resemble any typical emission-line objects, such as planetary nebulae,
symbiotic stars or T~Tauri stars.  Its is also significantly different from
that displayed by V4332~Sgr during the 1994 eruption (\cite{martini}) when
the emission-line spectrum was dominated by the Balmer series.  To some extent
the present spectrum of V4332~Sgr can be compared to those of pecular
giants, such as U~Equ (e.g. Barnbaum et~al. \cite{barn}) or VY~CMa (e.g.
Herbig \cite{herbig}).

The strongest emission features in the spectrum of V4332~Sgr are identified
in Fig.~\ref{spectr_f}.  Table~\ref{spectr_t} presents details of the
emission lines, i.e. measured wavelengths (in \AA), line fluxes (in
$10^{-15}$~erg~cm$^{-2}$~s$^{-1}$) with errors in parenthesis, full
width half maxima (in \AA), equivalent widths (in \AA) and the proposed
identifications. For all lines the flux, FWHM and EW were derived by
performing a Gaussian fit to the observed profile using the REWIA package (J.
Borkowski; http://www.ncac.torun.pl/$^\sim$jubork). The errors on the line
fluxes mostly depend on the uncertain continuum level and have therefore been
calculated in an aproximate way as a product of the half of the average noise
level and a doubled FWHM of a given line.

\begin{table}
\centering
\caption{Emission line spectrum of V4332 Sgr observed on 3~April~2003.
 [$\lambda_\mathrm{obs}$, FWHM and EW are in \AA. 
 Flux is in $10^{-15}$~erg~cm$^{-2}$~s$^{-1}$.
 Identification gives $\lambda_\mathrm{lab}$ (\AA), element (molecule) and 
 transition (for molecular transitions 
 (v'-v") means the upper and lower vibrational levels).]}
\label{spectr_t}
\begin{tabular}{@{} r @{~~~} r @{} r @{~} r @{~~} l}
\hline
$\lambda_\mathrm{obs}$~~  &  Flux~~~~~ &  FWHM  &  EW  &
Identification \\
\hline
4606.7  &   7.6 ($\pm$3.3) &  17~~  &  82~~  &  \\
4846.9  &   7.8 ($\pm$2.0) &  11~~  &  52~~  &  4842.3  AlO $B\,^{2}\Sigma^{+}$\,$-$\,$X\,^{2}\Sigma^{+}$\,(0-0) \\
4866.7  &   6.8 ($\pm$2.4) &  12~~  &  44~~  &  4866.4  AlO $B\,^{2}\Sigma^{+}$\,$-$\,$X\,^{2}\Sigma^{+}$\,(1-1) \\
5078.1  &   8.4 ($\pm$2.3) &  12~~  &  41~~  &  5079.4  AlO $B\,^{2}\Sigma^{+}$\,$-$\,$X\,^{2}\Sigma^{+}$\,(0-1) \\
5105.9  &   9.4 ($\pm$2.6) &  14~~  &  44~~  &  5102.1  AlO $B\,^{2}\Sigma^{+}$\,$-$\,$X\,^{2}\Sigma^{+}$\,(1-2) \\
        &                  &        &        &  5110.4  FeI (1)\\
5123.1  &   3.3 ($\pm$1.7) &   9.2  &  15~~  &  5123.3  AlO $B\,^{2}\Sigma^{+}$\,$-$\,$X\,^{2}\Sigma^{+}$\,(2-3) \\
5164.3  &   4.8 ($\pm$1.4) &   7.4  &  21~~  &  5166.3  FeI (1)\\
        &                  &        &        &  5168.9  FeI (1)\\
5204.3  &  10.8 ($\pm$1.7) &   9.2  &  46~~  &  5204.6  FeI (1)\\
5294.4  &   3.8 ($\pm$1.2) &   6.3  &  15~~  &  \\
5802.2  &   7.4 ($\pm$3.9) &  21~~  &  19~~  &  \\
5891.0  &  64.5 ($\pm$2.2) &  12~~  & 155~~  &  5889.9  NaI \\
        &                  &        &        &  5895.9  NaI \\
6039.6  &  12.9 ($\pm$3.0) &  16~~  &  29~~  &  6036.1  ScO $A\,^{2}\Pi$\,$-$\,$X\,^{2}\Sigma$ (0-0) \\
6079.5  &  14.2 ($\pm$3.3) &  17~~  &  31~~  &  6079.2  ScO $A\,^{2}\Pi$\,$-$\,$X\,^{2}\Sigma$ (0-0) \\
6137.3  &   5.5 ($\pm$3.8) &  20~~  &  12~~  &  \\
6161.2  &  12.4 ($\pm$2.9) &  15~~  &  26~~  &  6161.6  TiO $\gamma'$ (0-0) \\
6190.7  &   6.5 ($\pm$2.4) &  13~~  &  14~~  &  6185.6  TiO $\gamma'$ (0-0) \\
6221.6  &   8.3 ($\pm$3.6) &  19~~  &  17~~  &  6216.8  TiO $\gamma'$ (0-0) \\
6403.5  &   9.0 ($\pm$4.1) &  22~~  &  18~~  &  \\
6569.8  &  15.7 ($\pm$1.6) &   8.4  &  32~~  &  6572.8  CaI ($^1$S$-^3$P$^0$) \\
6780.3  &   6.8 ($\pm$3.4) &  18~~  &  14~~  &  6781.3  TiO $\gamma$ (2-1) (?)\\
6840.3  &   9.3 ($\pm$6.5) &  34~~  &  19~~  &  6849.9  TiO $\gamma$ (3-2) (?)\\
        &                  &        &        &  6814.7  TiO $\gamma$ (3-2) (?)\\
6974.8  &  10.2 ($\pm$4.0) &  21~~  &  21~~  &  \\
7057.5  &  26.3 ($\pm$3.8) &  20~~  &  52~~  &  7054.3  TiO $\gamma$ (0-0)\\
7093.5  &  24.1 ($\pm$3.3) &  17~~  &  47~~  &  7087.7  TiO $\gamma$ (0-0)\\
7129.4  &  30.4 ($\pm$2.7) &  14~~  &  57~~  &  7125.5  TiO $\gamma$ (0-0)\\
        &                  &        &        &  7124.9  TiO $\gamma$ (1-1)\\
7160.9  &   4.7 ($\pm$2.2) &  11~~  &   8.6  &  7158.5  TiO $\gamma$ (1-1)\\
7199.4  &   5.5 ($\pm$2.4) &  13~~  &   9.5  &  7197.2  TiO $\gamma$ (1-1)\\
        &                  &        &        &  7197.2  TiO $\gamma$ (2-2)\\
7271.2  &   3.2 ($\pm$3.0) &  16~~  &   4.8  &  7270.4  TiO $\gamma$ (2-2)\\
7343.2  &   5.7 ($\pm$2.9) &  16~~  &   6.6  &  \\
\hline
\end{tabular}
\end{table}

The strongest emission feature in our spectrum of V4332~Sgr is due to the
unresolved Na\,I~D lines.  We have also been able to identify emission lines
from other neutral atoms, i.e. Fe\,I and Ca\,I (see below).

The most interesting aspect of the present spectrum of V4332~Sgr is that it
displays molecular bands in emission.  The strongest and most numerous
features are due to TiO.  We have identified the prominent emissions near
7100\,\AA\ with three components of band (0-0) of the $\gamma$ system
corresponding to the transitions between electronic levels $A\,^{3}\Phi -
X\,^{3}\Delta$.  We believe that the (1-1) band of the same system at
7100--7200\,\AA\ (not marked in Fig.~\ref{spectr_f}) is also visible in our
spectrum.  Another prominent band of TiO is that of (0-0) of the $\gamma'$
system ($B\,^{3}\Pi - X\,^{3}\Delta$) near 6200\,\AA.  Some other emission
features can also be identified with excited bands of the TiO~$\gamma$
system.  Following Herbig (\cite{herbig}) we have identified the prominent
emission features at 6040 and 6080\,\AA\ with ScO.  AlO, which dominates the
near-IR spectrum of \cite{baner}, is also present in emission in our
spectrum.  Four bands of the $B\,^{2}\Sigma^{+} - X\,^{2}\Sigma^{+}$
electronic transitions are clearly seen near 4850\,\AA\ and 5100\,\AA.

The strong feature at 6569.8\,\AA\ was initially attributed to
$\mbox{H}_\alpha$, but we subsequently identified it as the Ca\,I
intercombination transition.  $\mbox{H}_\alpha$ was seen in the spectrum of
V4332~Sgr during its 1994 eruption (\cite{martini}), but at that time its
mean position was 6558.7\,\AA.  Since our position of the Na\,I~D lines
differs only by 1.3\,\AA\ from that of \cite{martini}, it is difficult to
reconcile a difference of 11\,\AA\ in the $\mbox{H}_\alpha$ position.
\cite{banash} suggest that this feature might be due to
TiO~$\gamma'$~(0,1) as the wavelength matches well.  In that case, other
emissions originating from the same transition should be seen at
6596.3\,\AA\ and 6629.0\,\AA\, but they are not present.  Furthermore,
Table~\ref{spectr_t} shows that the 6569.8\,\AA\ feature is significantly
narrower than other molecular lines, suggesting that it is an atomic line. 
Given the very low excitation of the spectrum of V4332~Sgr, we consider
Ca\,I~6572.8\,\AA\ ($^1$S$-^3$P$^0$) as the most probable identification for
the 6569.8\,\AA\ feature.

As can be seen from Fig.~\ref{spectr_f}, the spectrum shows a clear
continuum rising toward longer wavelengths.  The rather low signal-to-noise
ratio does not allow unambiguous identification of absorption features.  Yet
a comparison of the observed spectrum with synthetic ones suggest an early M
type.

An analysis of the strengths and widths of molecular bands allows us to
estimate excitation conditions in the medium where the bands are produced.
For the
TiO bands we have used the data base described by Schwenke (\cite{schw}) and
distributed by Kurucz (\cite{kur}).  The observed widths of the emission
peaks of the (0-0) band of the $\gamma$ system of TiO (7050--7130\,\AA) has
allowed us to estimate the rotational temperature, $T_\mathrm{rot} \simeq
200$\,K.  On the other hand, the relative equivalent widths of these
emission peaks point to the electronic temperature, $T_\mathrm{el} \ga
500$\,K.  Following the analysis of Herbig (\cite{herbig}) and using the
observed equivalent-width ratio of the 6079\,\AA\ and 6036\,\AA\ peaks we
can estimate $T_\mathrm{rot} \simeq 800$\,K for ScO.  It is worth noting
that from numerous bands of AlO in the near-IR \cite{baner} estimated
$T_\mathrm{vib} \simeq 3000$~K and $T_\mathrm{rot} \simeq 300$~K. On the
other hand, from the water ice feature at 3.05~$\mu$m \cite{bva} have
estimated a temperature of 30$-$50~K, while their CO band analysis has
suggested 300$-$400~K.

\section{Analysis of the photometric data  \label{analys}}

Using the available photometric data we have attempted to analyse the
evolution of V4332~Sgr since its outburst in 1994.  The method is the same
as that used in Tylenda (\cite{tyl04}) in the analysis of V838~Mon. 
Basically it fits standard colours from a sequence of spectral types to the
photometric data.  From the best fit, obtained using the least squares
method, one derives the effective temperature, $T_\mathrm{eff}$, and the
angular stellar radius, $\theta$. This procedure requires the value of
interstellar extinction to be known. Finally, if the distance to the object is known,
the linear stellar radius, $R$, and the stellar luminosity, $L$, can then be
calculated.

\subsection{Interstellar extinction  \label{ext}}

The interstellar extinction to V4332~Sgr has been determined by \cite{martini}.
Comparing the photometric colours with the spectral classification from their spectra
obtained in March~1994 they have found $E_{B-V} = 0.32 \pm 0.02$. However,
the values of $E_{V-R}$ and $E_{V-I}$ they obtained were significantly bluer,
even negative in most instances (see their Fig.~6). This troublesome, at first
sight, inconsistency can, however, be easily explained. The intrinsic colours
which \cite{martini} compared with the photometric observations were in the Johnson
(\cite{johnson}) system while the observations of Gilmore were done in
the Cousins ($R_c$ and $I_c$) system (Gilmore, private communication). 
Indeed, as discussed in Sect.~\ref{analys_fr} and \ref{analys_ev}, 
there is a good agreement
between the spectral types of \cite{martini} and the spectral types resulting
from our fits to the $BVR_cI_c$ magnitudes if
$E_{B-V} = 0.32$ is assumed. We can also conclude
that the uncertainty of this value is not greater than 0.1. For
$E_{B-V}$ beyond this range our spectral classes from the photomertic data
become systematically discrepant from those of \cite{martini}. In particular,
from the fits done to the photometric data on 4--5~March (these data have been
used by \cite{martini} to compare with their spectrum on 4.5~March and to derive
$E_{B-V}=0.32$) assuming $E_{B-V} = 0.20$ we have obtained K4.5--6.5 while for
$E_{B-V} = 0.45$ the result is K2.0--3.0. This can be compared to K3--4 obtained
by \cite{martini} from their spectrum on 4.5~March.

In the present paper we assume $E_{B-V} = 0.32$.
The uncertainty in $E_{B-V}$ introduces uncertainties in the effective
temperatures and the angular stellar radius derived from the fits discussed below.
This errors can next propagate to
the values of the linear stellar radius and luminosity. We have found that an error of 0.1
in $E_{B-V}$ affects the effective temperature and the angular stellar
radius by at most 5\%. Thus in most cases 
this source of uncertainties is small compared to that due to 
the reference spectra (giants or supergiants)
used in the fitting procedure (see Table~\ref{evol_t}). The resulting error in luminosity
is larger, i.e. 15--20\%, but even larger uncertainties are expected to result from the distance
uncertainty (see Sect.~\ref{dist}).

\subsection{Progenitor and distance  \label{dist}}

MWT99  have estimated that the distance to V4332~Sgr might be 
$\sim$300~pc assuming that the object was a K giant at maximum.  They have
however concluded that this procedure did not lead them to consistent
results if later observations are considered.

Skiff (\cite{skiff}), using the POSS prints, identified a possible
progenitor of V4332~Sgr with $B$~=~18 and $R$~=~16. Using the ALADIN service
we have searched for brightness estimates of the V4332~Sgr progenitor in archival
surveys. We have found data in three catalogues. The SuperCOSMOS catalogue
for a star at the position of V4332~Sgr gives $B$~=~18.66, $R$~=~16.72
and $I$~=~15.89. In the USNO-B1.0 catalogue the figures are $B$~=~18.20,
$R$~=~16.48 and $I$~=~15.69.\footnote{Both, SuperCOSMOS and USNO-B1.0 give
second $R$ magnitudes for the V4332~Sgr progenitor, 
which are 14.96 and 14.38, respectively. From the $R$
magnitudes of three nearby stars in these catalogues which in the red images
are of similar brightness as the V4332~Sgr progenitor it can be safely concluded 
that these second magnitudes are significantly too bright.}
Finally in the USNO-A2.0 catalogue one finds
$B$~=~17.9 and $R$~=~16.4. The results are fairly consistent although 
there are systematic differences 
between the values in the three catalogues. In all the bands 
the object is faintest in SuperCOSMOS while brightest in USNO-A2.0. The
colours are more consistent. This is not surprising as in authomatic
measurements from survey plates or frames systematic errors are usually most
important.

With $E_{B-V} \simeq 0.32$ (see Sect.~\ref{ext}) the colours from the above
measurements imply an effective
temperature of 5000--6000~K.  For this temperature range, the most reasonable
assumption is that V4332~Sgr was a main sequence star prior to the eruption. 
Adopting intrinsic colours and absolute magnitudes for main sequence from
Drilling \& Landoldt (\cite{drill}) as well as assuming $E_{B-V} = 0.32$ and the 
standard extinction curve the magnitudes from SuperCOSMOS can be best fitted
with a K0\,V star at a distance of 1.35~kpc. The same procedure applied to the
results from USNO-B1.0 gives a G6\,V type and a distance of 1.58~kpc while
using the magnitudes from USNO-A2.0 one gets F8\,V and 2.6~kpc.

If one assumes that before eruption V4332~Sgr was a giant rather than a main
sequence star, then the observed colours of the progenitor would imply a
spectral type of G5--F8\,III and $M_V \simeq +1.0$ (Schmidt-Kaler \cite{sk}).
Then the distance would increase to $\sim$10--13~kpc. 
This solution seems, however, to be unlikely
given the position of V4332~Sgr in the sky, $l = 13\fdg63$ $b = -9\fdg40$,
and the observed reddening, $E_{B-V} \simeq 0.32$.  Several
planetary nebulae near this position, i.e. PNG 11.3--09.4,
12.5--09.8, 13.8--07.9, 13.7--10.6, 14.2--07.3, have $E_{B-V}$ in the range
0.37--0.52 (Tylenda et~al. \cite{task}), yet their distances are estimated
to be within 3--8.5~kpc (Acker et~al. \cite{aost}).

In the present paper we assume that V4332~Sgr was a solar type star
before eruption and that the object is at a distance of
$\sim$1.8~kpc (mean value from the above estimates adopting main sequence).
It is however clear from the above that this conclusion is subject 
to significant uncertainties. The distance value is probably uncertain to
50\% or so introducing corresponding uncertainties in the values of the stellar
radius and luminosity derived in Sect.~\ref{analys_ev}.

\subsection{Observational data and fitting them with standard spectra  
\label{analys_fr}}

During the 1994 eruption, the only multicolour ($UBVR_cI_c$) photometry of
V4332~Sgr was done by Gilmore (\cite{gilmore}) and the data cover the period
between 27~February and 11~March.  \cite{martini} have estimated the $BVR$
magnitudes from their spectra obtained on 11~March and 5--6~June~1994.  From
the published spectra in \cite{martini} we have made rough estimates of the
$I$ magnitude.  Their Fig.~4 shows that on 11~March the flux in the $I$ band
was $\sim$3.5 times stronger than in the $R$ band which results in $R-I
\simeq 2.2$.  From Fig.~5 in \cite{martini} one can estimate that on 
5--6~June the flux in the $I$ band was $\mathrm{F}_\lambda \simeq 3.5 \times
10^{-14}~\mathrm{ergs}~\mathrm{cm}^{-2}~\mathrm{s}^{-1}~\AA^{-1}$
giving $I \simeq 10.9$.

In the 2MASS survey the object was observed on 18~May~1998 and the results
were: $J=12.10$, $H=11.60$ and $K=10.99$.  V4332~Sgr was also measured in
the DENIS survey on 11~September~1999 which gave $I_{Gunn}=14.37$, $J=12.46$ and
$K_s=10.66$.  

In 2003, apart from our $BVR_cI_c$ photometry obtained on 21~May and 
summarized in Table~\ref{phot_t}, optical photometry was also done by 
\cite{banash} on 29~September. Besides,  $JHK$ magnitudes were obtained
by \cite{baner} on 19~June~2003, while $LM$ magnitudes measured on
5~September~2003 can be found in \cite{bva}. 
As discussed in Sect.~\ref{observ}, V4332~Sgr did not evolve significantly
between May and September. Therefore we have combined all the data from 2003
into one set for the analysis discussed below. In particular, in the optical
we have taken mean values from our results in Table~\ref{phot_t} and those
of \cite{banash}.

When fitting the observed magnitudes, we applied the standard intrinsic
colours for the luminosity class III (giants) and I (supergiants).  This is
justified for the period of the 1994 eruption as the observed spectra of
V4332~Sgr then resembled those of the III--I luminosity class
(\cite{martini}).  Also, for the data obtained in 2003 we have found that
the giant or supergiant spectra better fit the observations than the colours
for main sequence stars.  The intrinsic photometric colours have been
taken from Schmidt-Kaler (\cite{sk} -- $UBV$), Johnson (\cite{johnson}
-- $VRI$ for G and K types), Lee (\cite{lee} -- $VRI$ for M types),
Bessell \& Brett (\cite{bess} -- $VI_cJHKLM$ for giants) and Koornneef
(\cite{koor} -- $VJHKLM$ for supergiants).  The calibration of the effective
temperature and the bolometric correction against the spectral type has been
taken from Schmidt-Kaler (\cite{sk}).

\begin{figure}
\centering
  \includegraphics[width=8.0cm]{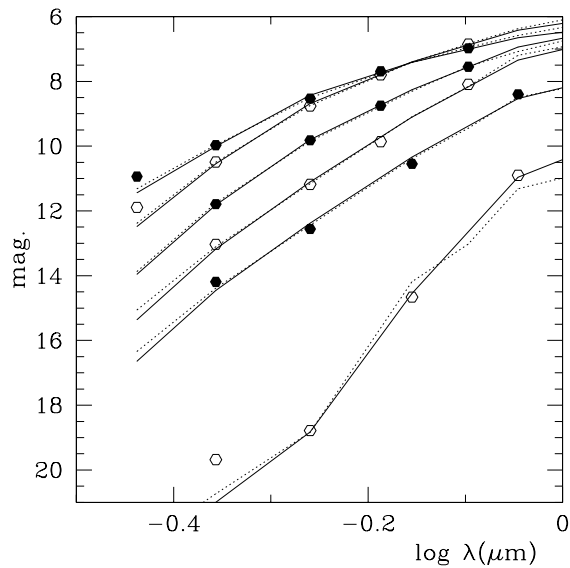}
  \caption{Evolution of the spectrum (magnitude versus 
effective wavelength) of V4332~Sgr in 1994.  Symbols
show the observed magnitudes (taken from Gilmore \cite{gilmore} 
and \cite{martini}).  Solid and dotted curves show the fitted standard 
supergiant and giant spectra, respectively (reddened using the standard 
extinction curve and $E_{B-V} = 0.32$).  From top to bottom the curves 
correspond to the observations made on 27.7~February, 4.7~March, 7.7~March,
10.7~March, 11.4~March and 5--6 June 1994, respectively.}
  \label{fit94}
\end{figure}

Figure \ref{fit94} presents the best fits of the standard giant (dotted
curves) and supergiant (solid curves) spectra to the photometric data
obtained in 1994.  Note that for clarity of the figure the spectra on
3.7~March and 8.7~March have not been displayed in Fig.~\ref{fit94}.

The results of the fits, i.e. spectral types, effective temperatures and
angular stellar radii, $\theta$ (in radians), are given in
Table~\ref{evol_t}. For each
date the results of the fits are given in two lines: first the results from
fitting the supergiant (I) spectra, second those for -- the giants (III).  
The lines marked with M in the fourth column give the results from
\cite{martini}, i.e. spectral type from classification and effective
temperature from stellar atmosphere model analysis of their spectra.

Note that the $U$ magnitude, measured by Gilmore only for three initial
dates, has not been taken into account in the fitting procedure as its
inclusion degrads the fit quality.  This is not suprising as the $U$ band is
dominated by the Balmer continuum which is very sensitive to nonstandard
phenomena such as departures from hydrostatic equilibrium, non-LTE and
winds.  As can be seen from Fig.~\ref{fit94}, V4332~Sgr was systematically
brighter by $\sim$0.5~magnitude in the $U$ band when compared to the
standard supergiants and giants.

Note also that when fitting the magnitudes derived from \cite{martini} for
5--6~June, the $B$ magnitude was not taken into account.  It was presumably
affected by strong emission lines, as can be seen from Fig.~5 in
\cite{martini}. The results obtained for this date are particularly uncertain
and should be regarded with caution. They are based on rough estimates
of the $VRI$ magnitudes made from the spectral observations. Besides, judging
from the estimated effective temperature the observations covered only
the shortwavelength Wien's part of the spectrum.
The bulk of the energy of the object was presumably emitted in 
the infrared where no measurements were made.
Finally, our fits for this date result from extrapolation
beyond the range of the standard spectra (the latest types for which intrinsic
colours are available are: M6--7 for giants and M5--6 for supergiants).

\begin{figure}
  \includegraphics[width=8.0cm]{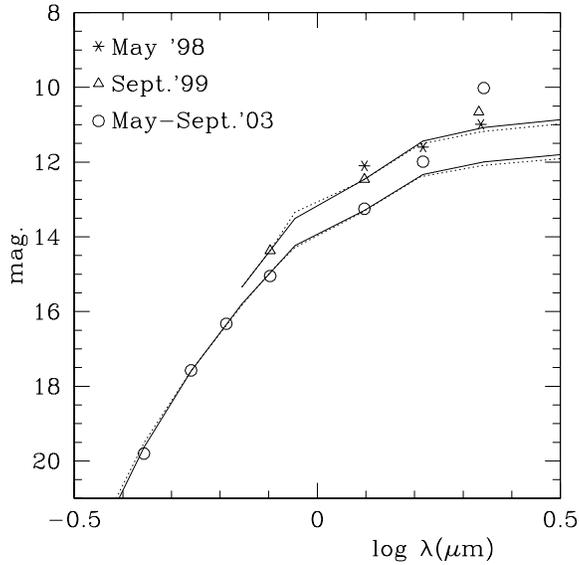}
  \caption{The spectrum of V4332~Sgr in 1998, 1999 and 2003. 
Asterisks -- $JHK$ in May~1998 from 2MASS.
Triangles -- $I_{Gunn}JK_s$ in September~1999 from DENIS.
Circles -- $BVR_cI_cJHK$ on May--September~2003 (see the text). 
Full and dotted curves -- standard supergiant and giant (respectively) spectra 
fitted to the 1999 and 2003 data (for details see the text).
Parameters of the fits are given in Table~\ref{evol_t}.
The model spectra have been reddened with $E_{B-V} = 0.32$.}
  \label{sp03}
\end{figure}

Figure \ref{sp03} shows the results of observations in 1998, 1999 and 2003. 
Asterisks indicate the 2MASS measurements obtained on 18~May~1998. 
Triangles show the DENIS results derived on 11~September~1999.
Circles represent the spectrum observed in 
May-September~2003 (see above for the sources fo the data).

First we discuss the observations obtained in 2003 as they cover the largest
spectral range.
The whole spectrum, i.e. all the circles in Fig.~\ref{sp03},
cannot be fitted with a single standard spectrum.  This can only be done for
shorter wavelengths. In Fig.~\ref{sp03} we show the best fits of the supergiant
(full curve) and giant (dotted curve) spectra for the $VR_cI_cJ$ measurements. 
The parameters of the fits are given in Table~\ref{evol_t}.  The $B$
magnitude has not been taken into account in the fitting procedure because
of its significant uncertainty.  However, as can be
seen from Fig.~\ref{sp03}, it fits well the obtained spectra. In the
long wavlength range the $H$ and, particularly, $K$ magnitudes show a clear
excess compared to the spectra fitted in the shorter wavelengths. 
With the $L$ and $M$ magnitudes measured by \cite{bva} 
(not shown in Fig.~\ref{sp03}) one can easily conclude 
that the source of this excess dominates the brightness of the object in the
infrared. This infrared excess will be discussed in Sect.~\ref{ire}.

The infrared observations displayed in Fig.~\ref{sp03} show that V4332~Sgr
evolved systematically between 1998 and 2003. It was becoming fainter in $J$
(and $I$) but brighter in $K$.  This can be interpreted
as due to a gradual fading
of the main object (seen in $J$ and $I$) and the increasing infrared excess
in $K$.  Therefore when fitting the standard spectra to the DENIS magnitudes
obtained on 11~September~1999 (triangles in Fig.~\ref{sp03}), expecting that
$K$ may be affected by the IR excess, we considered only $I$ and $J$. 
Indeed, as can be seen from Fig.~\ref{sp03}, the $K$ magnitude is $\sim$0.5
magnitude above the fitted spectrum. Note, however, that if all the three
($IJK$) measurements are taken into account the fits would be quite poor and
$T_\mathrm{eff}$ would lower by $\sim$100\,K for class III and $\sim$200\,K 
for class I compared to the values given in Table~\ref{evol_t}.

No fit to the 2MASS data on 18~May~1998 (asterisks) is shown in
Fig.~\ref{sp03}.  The spectral range of the data is quite narrow and hence
any $T_\mathrm{eff}$ between $\sim$3000~K and $\sim$4500~K could be
considered to satisfactorily fit the data.  Fortunately for this wide
$T_\mathrm{eff}$ range, log~$\theta$ of the fits varies by less than 0.20. 
The mean values of log~$\theta$ from these fits are given in
Table~\ref{evol_t}.

\subsection{Evolution of the object  \label{analys_ev}}

The last two columns in Table~\ref{evol_t} show the stellar radius and
luminosity (given in solar units) of V4332~Sgr calculated from the
angular radius ($\theta$) and the effective temperature ($T_\mathrm{eff}$)
assuming a distance of 1.8~kpc.

\begin{table}
\centering
\caption{Evolution of V4332 Sgr in 1994--2003}
\label{evol_t}
\begin{tabular}{c c c c c c}
\hline
 ~~~~Date& ~~~~~~~Sp.type & ~~~~~~~$T_\mathrm{eff}$ & ~~$-$log $\theta$ 
& $R/R_{\sun}$ & $L/L_{\sun}$ \\
\end{tabular}
\begin{tabular}{@{~} r l l r @{.} l r @{.} l r @{.} l @{~}}
\hline
27.7 Feb.'94 & K0.2 I     &  4400. &  8&85 &114&  &4360&  \\
             & K1.7 III   &  4470. &  8&86 &110&  &4380&  \\
 3.7 Mar.'94 & K3.3 I     &  4050. &  8&75 &142&  &4880&  \\
             & K3.6 III   &  4070. &  8&75 &141&  &4950&  \\
 4.5 Mar.'94 & K3-4 III-I &  4400. & M\\
 4.7 Mar.'94 & K4.0 I     &  3950. &  8&73 &150&  &4910&  \\
             & K4.2 III   &  3970. &  8&73 &149&  &4980&  \\
 7.7 Mar.'94 & M1.1 I     &  3640. &  8&75 &141&  &2790&  \\
             & M1.1 III   &  3710. &  8&80 &128&  &2790&  \\
 8.7 Mar.'94 & M2.4 I     &  3340. &  8&71 &154&  &2690&  \\
             & M2.7 III   &  3560. &  8&78 &133&  &2550&  \\
 9.5 Mar.'94 & M3   III-I &  3800. & M\\
10.7 Mar.'94 & M3.6 I     &  3070. &  8&67 &169&  &2290&  \\
             & M4.1 III   &  3420. &  8&82 &119&  &1760&  \\
11.4 Mar.'94 & M5-6 III-I &  3100. & M\\
             & M3.8 I     &  3040. &  8&89 &102&  & 790&  \\
             & M4.1 III   &  3420. &  9&08 & 66&  & 540&  \\
20.4 Mar.'94 & M6-7 III-II&  2600. & M\\
5--6 Jun.'94 & M8-9 III   &  2300. & M\\
             & M9.0 I     &  2120. &  8&97 & 86&  & 135&  \\
             & M8.0 III   &  3050. &  9&87 & 10&7 &   8&9 \\
  18 May '98 &            &        &  9&96 &  8&7 \\
             &            &        & 10&06 &  6&9 \\
  11 Sep.'99 & M3.9 I     &  3000. &  9&88 & 10&6 &   8&2 \\
             & M4.4 III   &  3390. & 10&06 &  6&9 &   5&6 \\
May-Sept.'03 & M2.7 I     &  3280. & 10&13 &  5&9 &   3&6 \\
             & M2.9 III   &  3540. & 10&22 &  4&8 &   3&2 \\
\hline
\end{tabular}
\end{table}

\begin{figure*}
\centering
  \includegraphics[width=5.9cm]{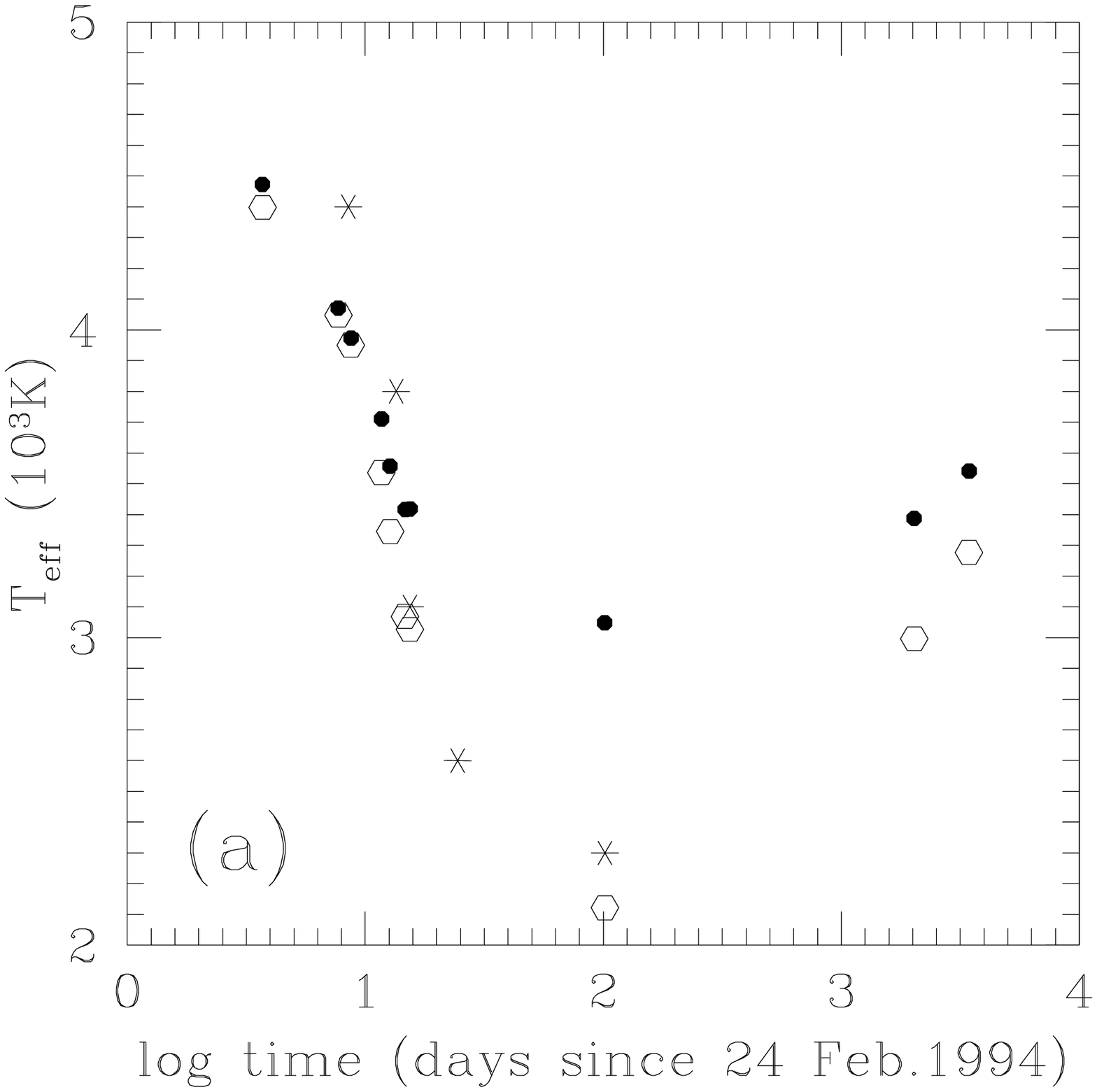}
  \includegraphics[width=5.9cm]{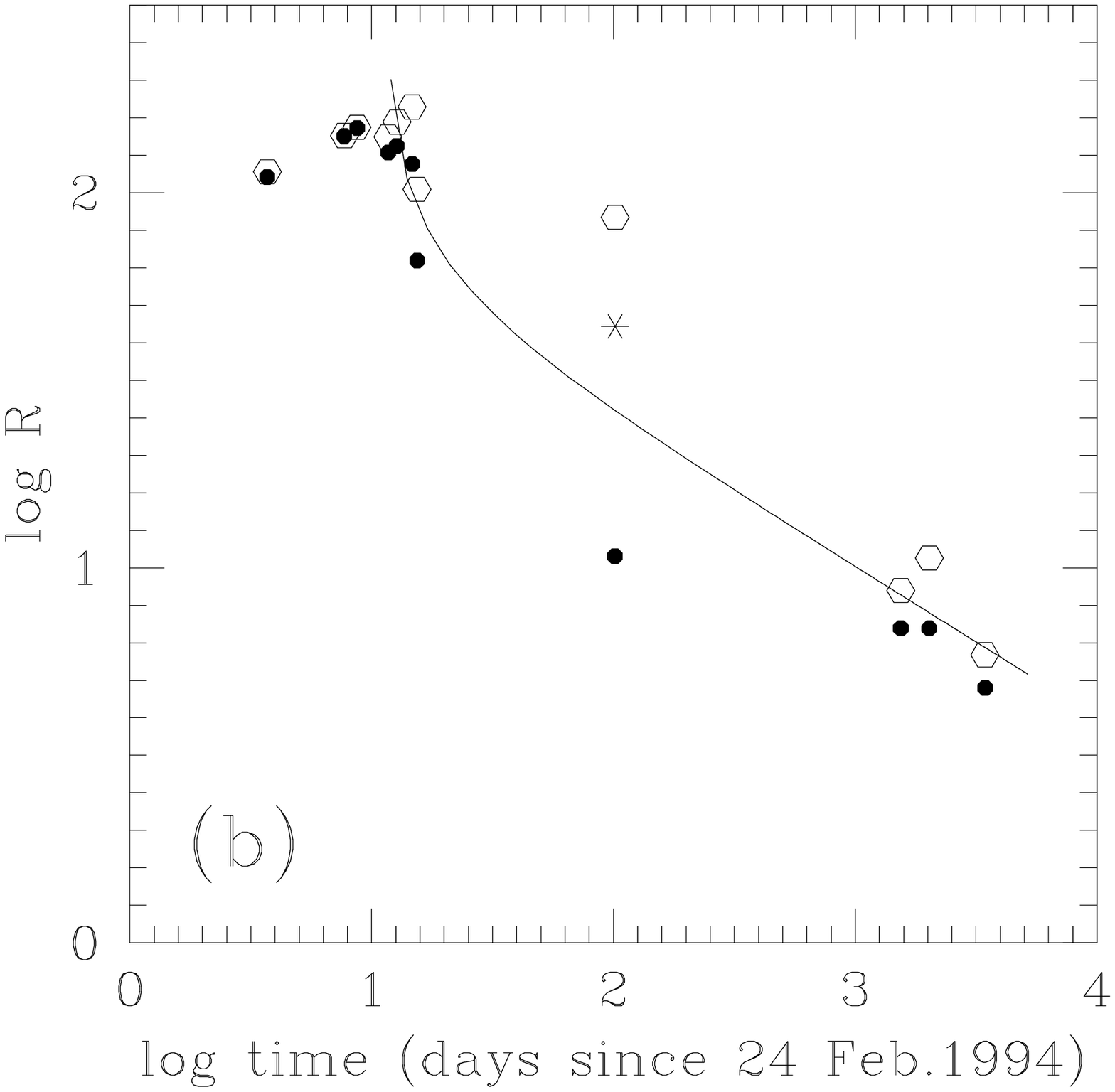}
  \includegraphics[width=5.9cm]{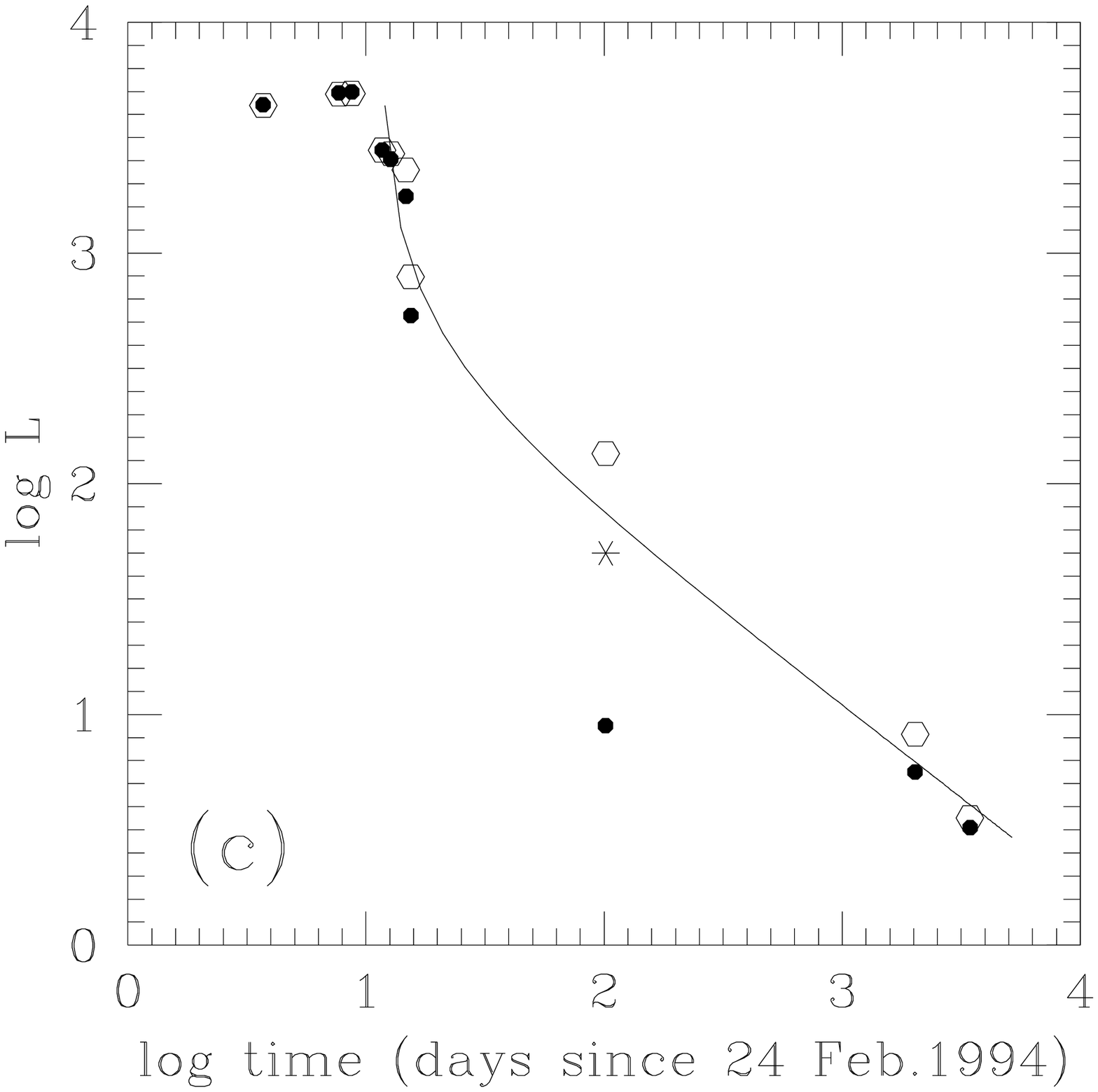}
  \caption{Evolution of V4332~Sgr with time.
Time (displayed in log scale) is in days counted since the discovery of 
the object in eruption (24~Feb.~1994). \emph{Left (a)} -- effective temperature in $10^3$~K. 
\emph{Centre (b)} -- logarithm of radius in $R_{\sun}$. \emph{Right (c)} -- logarithm of
luminosity in $L_{\sun}$. Open symbols -- results from fitting the
supergiant (I) spectra. Full points -- results from fitting the
giant (III) spectra. Asterisks in \emph{(a)} -- effective
temperature from \cite{martini}. Asterisks in \emph{(b)} and \emph{(c)} --
estimates on 5--6~June~1994 based on the results of \cite{martini} as
discribed in the text. Full curves in \emph{(b)} and \emph{(c)} --
results of a model discribed in Sect.~\ref{contrac}.
}
  \label{evol_f}
\end{figure*}

Figure \ref{evol_f} displays the evolution of principal parameters of
V4332~Sgr taken from Table~\ref{evol_t}.  The time is in days since 24
February 1994, i.e. since the discovery of V4332~Sgr in eruption.  The
effective temperature is in units of $10^3$~K, while the stellar radius and
luminosity are in solar units.  Open symbols show the results from fitting
the standard supergiant (I) spectra to the observations.  The same but using
the giant (III) spectra is presented with the full points.  Asterisks in
panel \emph{(a)} denote the effective temperatures derived by \cite{martini}
from fitting stellar atmosphere models to their spectra.

Let us first discuss the spectral types obtained from our analysis.  As can
be seen from Table~\ref{evol_t} no matter which luminosity class of the
standard spectra (I or III) is used, the resultant spectral types are always
very similar.  Moreover, our spectral classes are also very close to those
obtained by \cite{martini} from the classification of their spectra.  This
shows that the general approach adopted in our analysis of the photometric
data is consistent and reliable.

The consistency of the effective temperatures derived in the different ways
is not as good as that of the spectral types.  As can be seen from
Table~\ref{evol_t} and Fig.~\ref{evol_f}\emph{(a)} initially, when the
object was of the K type, both kinds of standard spectra give practically
the same values of $T_\mathrm{eff}$.  Later, when the object was becoming
cooler, an increasing discrepancy appears in the sense that $T_\mathrm{eff}$
obtained using the supergiant spectra is systematically lower than that from
using the giant spectra. The reason lies in the calibrations of
$T_\mathrm{eff}$ versus spectral type used in our study (Schmidt-Kaler
\cite{sk}). For the same spectral type, the value of $T_\mathrm{eff}$ for
supergiants is systematically lower than that for the giants but for the G,
K types the difference is small.  For the M types the discrepancy increases
and becomes as large as 530~K at M5.  This has obvious consequences for the
derived effective radii and luminosities: the use of the supergiant
standard spectra usually results in larger radii and luminosities.

Comparing our results with those obtained by \cite{martini}, one sees that
on March~4 and 9 their $T_\mathrm{eff}$ is systematically higher.  This
agrees with what was noted in \cite{martini}, namely that
the value of $T_\mathrm{eff}$ they obtained from the stellar model
analysis was too high for the observed spectral type.  \cite{martini}
ascribed this discrepancy to the narrow observed spectral range.  On the
later dates the spectral types and $T_\mathrm{eff}$ derived in
\cite{martini} are consistent and, as can be seen from Fig.~\ref{evol_f}a,
their $T_\mathrm{eff}$ follows quite closely our values obtained using the
supergiant spectra.

Note that, as discussed in Sect.~\ref{analys_fr}, our results obtained from
the data on 5--6~June~1994 (symbols at log~$t \simeq 2.0$ in
Fig.~\ref{evol_f}) are subject to significant uncertainty. They are
based on uncertain observational estimates done only in the Wien's part of 
the spectrum, as well as, on the extrapolated standard spectra.  
Due to the increasing
discrepancy in the effective temperature between giants and supergiants for
the late M types, discussed above, the difference between our estimates of
$T_\mathrm{eff}$ is as large as 900~K.  This resulted in large differences in
the stellar radius and luminosity seen in Table~\ref{evol_t} and
Fig.~\ref{evol_f}. \cite{martini}, from their model atmosphere analysis,
estimated that by 5--6~June the object had faded by a factor of 100 in
luminosity compared to the beginning of March.  Adopting our determinations
of $\sim$5000\,$L_{\sun}$ for the beginning of March this results in an
estimate of $\sim$50\,$L_{\sun}$ and, assuming
$T_\mathrm{eff} = 2300$\,K (from \cite{martini}), an effective radius of
$\sim$44\,$R_{\sun}$ on 5--6~June. These estimates are shown with asterisks in
Fig.~\ref{evol_f}\emph{(b)(c)}.

As can be seen from Table~\ref{evol_t} and Fig.~\ref{evol_f}c during
$\sim$10 days after its discovery, V4332~Sgr sustained a fairly constant
luminosity, the star being $\sim$5000 times brighter than before the
eruption.  The radius was then $\sim$140 times larger than that of the
progenitor, possibly slowly expanding with time.  Near 5~March, the stellar
luminosity declined rapidly and during $\sim$6~days the object faded by a
factor $\sim$7.

Concerning the effective radius we can only state, given the uncertainties
in our results, that the radius was not changing significantly till
10~March.  It seems, however, that a significant shrinkage of the radius
occurred near 11~March.  The initial luminosity decline was thus primarily
due to the decline in the effective temperature.

The subsequent fading involved the decline of all the stellar parameters, i.e.
$L$, $T_\mathrm{eff}$ and $R$.  At a certain time $T_\mathrm{eff}$ reached a
minimum but the available data do not permit an estimate of when it occurred
and how deep this minimum was.  From Fig.~\ref{evol_f} it can be inferred
that this occured near June~1994 or later.  The lowest observed
$T_\mathrm{eff}$ was reached on 5--6~June~1994, i.e. some 90 days after the
luminosity started declining.  During this time interval, the object dropped
in luminosity by factor at least $\sim$40.  According to \cite{martini},
this luminosity drop was of factor 100 and the object became as cool as
2300~K.

The decline in luminosity and radius has continued, although at a decreasing
rate.  This has been followed by a slow rise in the effective temperature
and at present the object is significantly hotter than in June~1994. 
Comparing the results obtained in 2003 with the estimates made for the
progenitor (see Sect.~\ref{dist}), 
we can conclude that V4332~Sgr is now larger and more luminous,
but cooler than before the eruption, so it has not yet returned to its
pre-outburst state.

\section{Gravitational contraction in the decline  \label{contrac}}

The fading of V4332~Sgr can be consistently interpreted if one assumes that
the energy source, that caused V4332~Sgr to erupt to $\sim$140~$R_{\sun}$,
suddenly disappeared.  This resulted in a gravitational contraction of the
inflated envelope and a release of its gravitational energy.  To show
this, let us assume that the inflated envelope has a density distribution of
$\rho \sim r^{-5/2}$ (following Soker \& Tylenda \cite{soktyl}).  Then,
assuming that the envelope mass, $M_\mathrm{e}$, is small compared to the star mass,
$M_\star$, it is straightfoward to show that the gravitational energy of the
envelope is 
\begin{equation}
  \label{eg}
E_\mathrm{g} = -\frac{G M_\star M_\mathrm{e}}{(R_\star R)^{1/2}},
\end{equation}
where $G$ is the gravitational constant, $R_\star$ is the radius of the
unperturbed stellar layers (assumed to be the inner radius of the envelope) 
and $R$ is the outer radius of the envelope.  If the radiated luminosity,
$L$, is due to gravitational contraction of the envelope, one gets
\begin{equation}
  \label{l1}
L = -\frac{\mathrm{d}E_\mathrm{g}}{\mathrm{d}t} = 
   \frac{G M_\star M_\mathrm{e}}{2 R_\star^{1/2}}
   \frac{1}{R^{3/2}} \frac{\mathrm{d}R}{\mathrm{d}t}.
\end{equation}
On the other hand 
\begin{equation}
  \label{l2}
L = 4 \pi R^2 \sigma T_\mathrm{eff}^4.
\end{equation}
Combining Eq.~(\ref{l1}) with Eq.~(\ref{l2}) gives
\begin{equation}
  \label{dr}
\frac{\mathrm{d}R}{R^{7/2}} =
  -\frac{8 \pi R_\star^{1/2} \sigma T_\mathrm{eff}^4}{G M_\star M_\mathrm{e}} 
   \mathrm{d}t.
\end{equation}
Assuming that $T_\mathrm{eff}$ is constant (which is obviously a very crude
approximation), Eq.~(\ref{dr}) can be integrated giving
\begin{equation}
  \label{r}
\frac{1}{R^{5/2}} = \frac{20 \pi R_\star^{1/2} \sigma T_\mathrm{eff}^4}
  {G M_\star M_\mathrm{e}} t + \frac{1}{R_0^{5/2}},
\end{equation}
where $R_0$ is the initial envelope radius assumed to start contracting at
$t = 0$. 

The solid curves in Fig.~\ref{evol_f}\emph{(b)(c)} shows the evolution of a
gravitationally contracting envelope calculated from Eq.~(\ref{r}). 
The solar values for the stellar paramaters, i.e. $M_\star = 1.0 M_{\sun}$ and 
$R_\star = 1.0 R_{\sun}$, have been assumed, as well as, 
$T_\mathrm{eff} = 3300$\,K as a typical effective temperature during the
decline.  The mass of the envelope have been chosen to fit the observed
evolution and in the case shown in Fig.~\ref{evol_f}\emph{(b)(c)},
$M_\mathrm{e}$ is $1.5\times 10^{-5}M_{\sun}$.  We note that values of
$M_\mathrm{e}$ differing by less than a factor 3, i.e. being within 
$0.5-5.0\times10^{-5}M_{\sun}$, can be considered as reasonably reproducing the
observations.  Given the crudness of our approach and uncertainties in the
observational determinations of $R$, we can conclude that pure gravitational
contraction of the envelope inflated during the eruption can satisfatorily
explain the observed decline of V4332~Sgr.  The mass of the envelope is
small, of order $10^{-5} M_{\sun}$.  Note that the total mass involved in
the eruption was probably significantly larger than $M_\mathrm{e}$.  Some
mass was probably lost from the object and/or is circulating the object in a
disc-like structure (see Sect.~\ref{ire}).

\section{Infrared excess in 2003  \label{ire}}

One of the most important features in the spectrum of V4332~Sgr in 2003 is
the infrared excess clearly seen in the $H$ and $K$ bands in
Fig.~\ref{sp03}.  As discussed in Sect.~\ref{analys_fr}, no such excess is
necessary to explain the data in 1998.  The 2MASS magnitudes can be well
accounted for by a cool star somewhat brighter then the stellar component
seen in 2003.  In September~1999 the $K$ band was probably affected by an 
excess, although its magnitude was certainly significantly smaller than 
that in 2003.

\begin{figure}
  \includegraphics[width=8.0cm]{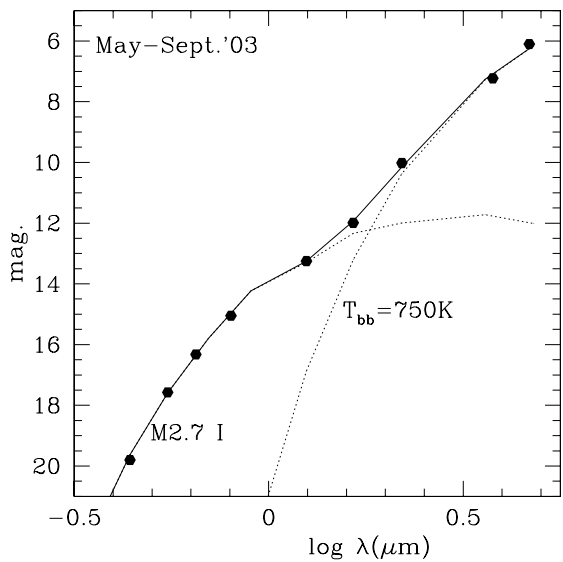}
  \caption{The magnitudes observed in May--September~2003 (see
Sect.~\ref{analys_fr} for the sources of the data)
fitted with a two component spectrum: M2.7 supergiant of 
$T_\mathrm{eff} = 3280$~K and blackbody of $T = 750$~K. Symbols -- observed
magnitudes. Dotted curves -- individual component spectra. Full curve -- sum of
the two components.}
  \label{apr_f}
\end{figure}

In order to discuss the origin of this IR excess, we have attempted to fit
the observations in 2003 with multi-component spectra.
Symbols in Fig.~\ref{apr_f} show the spectrum in May-September~2003
(see Sect.~\ref{analys_fr} for the sources of the data).
The full curve shows a fit of the data
with a two component spectrum: a standard supergiant spectrum of
$T_\mathrm{eff} = 3280$~K (M2.7~I -- the same as in Fig.~\ref{sp03})
and a blackbody distribution with $T_\mathrm{bb} = 750$~K.  As can be seen
from Fig.~\ref{apr_f} the fit is very good.
The angular radius of the blackbody component in Fig.~\ref{apr_f} is
log~$\theta = -8.28$, which at a distance of 1.8~kpc gives an effective
radius of $\sim$400\,$R_{\sun}$ and a luminosity of
$\sim$50\,$L_{\sun}$.

Fits of the kind shown in Fig.~\ref{apr_f} can be interpreted as evidence
for a dusty envelope that reprocesses radiation absorbed from the central
star. This interpretation is however very unlikely.
The infrared component is $\sim$15 times brighter than our estimate of
the stellar luminosity in Table~\ref{evol_t}.  This would imply that the
central star is significantly more luminous and that what we observe in the
optical is the stellar spectrum severely attenuated by dust.
Then the central star, reddened by dust, would be much hotter than 
estimated from the photometry.  Yet,
as noted in Sect.~\ref{spectr}, our spectrum suggests an early M type star, thus
consistent with the photometric results.  Besides, the large present
luminosity and the presence of hot nearby dust would inevitably indicate
that V4332~Sgr has recently experienced another outburst.
With the effective radius
of the blackbody component estimated above ($\sim$400\,$R_{\sun}$) and
adopting a rather lower limit of 10~km/s for the expansion velocity, one finds that
dust would have been lost a year before May-September~2003. However, as discussed in 
Sect.~\ref{analys_fr}, the infrared excess started developing in 1999.
Besides, there is no other observational indication that the object recently 
suffered from a secondary outburst. On the contrary, as can be seen from
Fig.~\ref{evol_f}, the object has been following the long
term decline after the 1994 eruption. The lack
of any significant evolution in the optical magnitudes between May and 
September~2003 is also an evidence for the long time scale evolution of the
object in 2003.

BVA04 have recently considered that V4332~Sgr might be a young object
surrounded by a protostellar circumstellar disc. This hypothesis has been invoked
to explain the origin of the water-ice band observed in the infrared spectrum. 
It also allows
\cite{bva} to suggest, following the ideas of Soker \& Tylenda (\cite{soktyl})
and Retter \& Marom (\cite{retmar}), that the 1994 outburst of V4332~Sgr might 
have been due to infall of an inner planet due to tidal interaction with the disc.
According to \cite{bva} the presently observed emission spectrum and
the infrared excess, which appeared in 1999, might have been due to interaction
of the 1994 ejecta with the circumstellar disc. However, an interaction of 
this kind should have happen during the 1994 eruption or soon after it. 
Assuming an expansion velocity of 100~km/s (\cite{martini}) 
the distance of 400~$R_{\sun}$ is reached in a month. Thus it is not clear
why the IR excess might have appeared $\sim$5 years later and why it might 
have been brightening with time, as observed.

The idea of a protostellar disc can however be used to explain the increasing
IR excess, although in a different way to that in \cite{bva}. In the case of 
an inner planet and an external accretion disc, tidal forces transfer angular
momentum from the orbital motion of the planet to the inner edge of the disc. 
As a result the inner edge of the disc
is kept at a certain distance from the planet orbit, determined by a balance 
between the rate of the angular momentum gain from the planet and the rate
in which the angular momentum is transported outward in the disc. When the 
planet gets accreted by the central star the eruption can disturb the disc,
possibly destroying its inner regions. After the event
the inner disc does not gain angular momentum any more.
As a result the inner edge will be approaching 
the star and the inner disc will become hotter and hotter. Thus we can
speculate that just after the 1994 eruption the inner disc in the V4332~Sgr system 
was far away, cold and thus radiating only in the far IR, while at present 
the inner disc regions are closer to the star, thus they are hotter and 
dominating the observed brightness in the $KLM$ bands.

There are, however, observational facts which do not favour the hypothesis
that V4332~Sgr is a young object.
First, the position of V4332~Sgr in galactic coordinates is $l = 13\fdg63$,
$b = -9\fdg40$.
According to catalogues of Sharpless (\cite{sharp}), Lynds (\cite{lynds}), 
Avedisova (\cite{aved}) and Russeil (\cite{russ}) there is no HII or 
star forming region closer than $5\degr$ from the position of V4332~Sgr.
Also from the CO map of Dame et~al. (\cite{dame}, Fig~2) one can see that
the object lies well outside any significant CO emission. 
Second, \cite{martini}, from their high 
resolution spectra near H$_\alpha$, have derived a radial velocity
of V4332~Sgr to be $-$180~km/s. A similar value, i.e, $-$160~km/s, can be obtained
from the observed positions of all the emission lines listed in their Table~3. 
This result has been derived from outburst spectra so interpretations other than 
the radial velocity of the object (although very unlikely) can be considered 
(see \cite{martini}). However, 
the observed wavelengths of the CaI line in our Table~\ref{spectr_t}
still gives a similar radial velocity, i.e $-$140~km/s.
The Galactic rotation curve (see e.g. Brand \& Blitz \cite{bb93}) predicts,
for the position of V4332~Sgr and a distance of 1.8~kpc, $V_{LSR} = +13$~km/s
which is equivalent to a heliocentric radial velocity of $\sim +3$~km/s.
A CO radial velocity map of Dame at~al. (\cite{dame}, Fig.~3) shows, at 
the galactic longitude of V4332~Sgr, $V_{LSR}$
between 0 and +140~km/s. Thus V4332~Sgr
does not follow the Galactic rotation which is not expected to be the case 
for a young object. 

Formation of a circumstellar disc is, however, also possible in
the merger scenario proposed by Soker \& Tylenda (\cite{soktyl}) in which
a star accretes a less massive object.
In this case, the inflated stellar envelope gains
not only energy from the merger, but also angular momentum from the orbital
motion. Note that if the merger event is provoked by an interaction with
a fast moving object (as discussed above V4332~Sgr is probably moving with 
a velocity $\ga$200~km/s relative to the objects rotating in the Galactic disc)
the angular momentum accreted is likely to be much larger than that of 
the Keplerian motion near the star surface.
During the outburst, i.e. when the envelope remains inflated to
large radii, processes due to turbulent or convective motions, tidal
effects, as well as magnetic fields are likely to transport angular momentum
outward.  When the merger process is more or less complete, the inflated
envelope would tend to collapse but the angular momentum stored in the
equatorial regions may prevent these regions from contracting significantly. 
They would form a ring-like or disc-like structure orbiting the central
object.  During the fading phase when the contracting central object is
likely to rotate significantly faster than the matter orbiting at larger
distances, magnetic fields and tidal forces can continue to transport
angular momentum outward.  With time, due to viscous processes, the
ring-like structure would evolve toward an accretion disc and dissipate its
energy.

Following the above discussion we have fitted the observations in 2003
with a composite spectrum of a central star and  
a standard blackbody accretion disc. The central star spectrum is the same
as in Fig.~\ref{apr_f}.
The distribution of the effective temperature
of the disc is (see e.g. Eq.~5.43 in Frank et~al. \cite{fkr})
\begin{equation}
\label{disc_t-eq}
  T(R) = T_0 \left\{ 3 \left( \frac{R_0}{R} \right)^3
         \left[1 - \left( \frac{R_0}{R} \right)^{1/2}
         \right] \right\}^{1/4}
\end{equation}
where
\begin{equation}
\label{t_0_eq}
  T_0 = \left( \frac{G M_\ast \dot{M}}{8 \pi R_0^3 \sigma} \right)^{1/4},
\end{equation}
$R_0$ is the inner radius of the disc while $\dot{M}$ is the accretion rate.
Note that $T(R)$ attains a maximum value $0.642\,T_0$ at $R = 1.36\,R_0$.
Integrating the blackbody spectrum with $T(R)$ over the disc surface one gets
the disc spectrum.

We do not show the final fit to the observations as it is very much the same 
as that in Fig.~\ref{apr_f}. It has been obtained for $T_0 = 1400$\,K and
log$(\theta_0) = -8.70$, where $\theta_0$ is the "observed" angular inner
radius of the disc. Assuming a distance of 1.8~kpc and $M_\ast = 1\,M_{\sun}$
this translates to $R_0 \simeq 160\,R_{\sun}$ and 
$\dot{M} \simeq 8.8 \times 10^{-4} M_{\sun}$/yr. The total luminosity of the disc
is then $\sim$87\,$L_{\sun}$. Note that $R_0$ is very close to the maximum
radius of the star during the 1994 eruption (see Table~\ref{evol_t}). This 
suggests that if the disc is of protostellar origin its inner regions have
probably been distroyed during the eruption, as discussed above.

Taking Eq~(5.69) in Frank et~al. (\cite{fkr}) 
and assuming the viscosity parameter, $\alpha = 0.01 - 0.1$, 
one gets, for the above values of $R_0$ and $\dot{M}$,
the viscous time scale of 4--20~years.  Thus one can argue that in 1998--1999 
the inner edge of the disc or ring-like structure was
somewhat farther away and cooler than in 2003, the star
was brighter so the relative contribution of the structure in the $HK$ bands
was negligible.

\section{Discussion and conclusions  \label{discuss}}

There is no doubt that the eruption of V4332~Sgr observed in 1994 was a
highly unusual, enigmatic and extremely interesting astrophysical event. 
This statement results not only from the observed behaviour of the object in
1994 but also from its present state marked by the unusual emission line
spectrum and the mysterious infrared excess. It is most unfortunate that so little
information is available regarding the evolution of V4332~Sgr during and
after its 1994 eruption.  The data is limited to the optical with just the
six multicolour photometric observations of Gilmore (\cite{gilmore})
covering 11 days and seven spectra of \cite{martini} obtained in a three
month period.  The lack of infrared data is particularly disappointing since
the object was likely emitting much of its energy in this wavelength region.

The results of our photometric analysis in Sect.~\ref{analys}, supplemented
with the results of \cite{martini}, clearly show that the decline of
V4332~Sgr following its 1994 eruption was marked by a decline in all
principal parameters: luminosity, radius and effective temperature.
In 2003 the object was $\sim$1500 times less luminous than during the eruption
but still remaining in the spectral range of the M type. As
discussed in Soker \& Tylenda (\cite{soktyl}), this type of evolution
rules out the classical nova mechanism which could have
worked for a hypothetical white dwarf companion to the progenitor. 
Note also that in June~1994 and May--September~2003, V4332~Sgr was significantly
(1.0--1.5~mag.) fainter in the $B$ band than the progenitor (between these
two dates there were no measurements in $B$).  This implies that it was
the progenitor which erupted in 1994 and not its hypothetical faint companion.

As discussed in Sect.~\ref{contrac}, the decline of V4332~Sgr can be
understood in terms of a pure gravitational contraction of the inflated
stellar envelope. An event of this kind is expected to follow 
a sudden switch-off of the energy sources which have so far been working at 
the base of the envelope.
This interpretation is well in line with the stellar merger scenario
of Soker \& Tylenda (\cite{soktyl}).  When the merger phase is complete, the
source of energy quickly drops and the inflated envelope has to contract under 
the gravity of the star. The considerations in Sect.~\ref{contrac} show 
that in order to explain the
observed decline of V4332~Sgr over almost a decade, the mass of the
inflated stellar envelope must be small, of order $10^{-5}M_{\sun}$. 

The origin of the infrared excess dominating the observed $KLM$ magnitudes
in 2003 is unclear. When fitted with blackbody it implies a temperature
of $\sim$750~K and a luminosity $\sim$15 times higher than that of the
central star. As discussed in Sect.~\ref{ire} it is very unlikely that
the excess is due to a dusty envelope reprocessing the radiation from 
the central star. We have considered that the IR radiation can originate from
a circumstellar disc dissipating its gravitational energy via viscous processes.
The disc can be either of protostellar origin, following suggestions of 
\cite{bva}, or being a remnant storing angular momentum after the 1994 eruption.
Both possibilities are in line with the merger scenario of Soker
\& Tylenda (\cite{soktyl}).
The first one assumes that V4332~Sgr is a young object. This hypothesis is however
inconsistent with the position of the object in Galaxy and, particularly, with
its radial velocity significantly different from the Galactic
rotation curve. Future observations of V4332~Sgr in the infrared, particularly
in longer wavelengths might be crucial for understanding the nature of the object.

A lower limit to the mass of this
object accreted in the merger scenario of Soker \& Tylenda (\cite{soktyl})
can be obtained from the total energy emitted in the eruption.  The
luminosity of V4332~Sgr integrated since the discovery of the 1994 eruption
till 2003 is $\sim 4.5 \times 10^{43}$\,ergs.  Equating this to
$G M_\star M_\mathrm{acc}/R_\star$ one obtains 
(assuming $M_\star = 1.0 M_{\sun}$ and $R_\star = 1.0 R_{\sun}$) 
$M_\mathrm{acc} \simeq 10^{-5} M_{\sun}$. 
Note that this estimate does not take into account the energy
radiated away before the discovery (it is very likely that the eruption of
V4332~Sgr started well before 28~February~1994), nor the energy stored
in the matter now circulating the central star, nor the kinetic
energy in mass loss. Given the above estimate as well as that of
$M_\mathrm{e}$ in Sect.~\ref{contrac}, we can conclude that the mass
of the accreted object was $\ga 10^{-4} M_{\sun}$.

The unusual emission line spectrum observed in April~2003 and described in
Sect.~\ref{spectr} deserves futher studies as it can provide important
insight into the current state of V4332~Sgr. From our
preliminary analysis in Sect.~\ref{spectr} we can conclude that it must
originate in an optically thin, neutral, molecular, cold medium.  Low
optical thickness means low column density which, with large equivalent
widths of many lines in the spectrum, requires large volume -- considerably
larger than that of the central star. This obviously rules out the
stellar atmosphere and regions in its near vicinity.  The low rotational
temperature inferred in Sect.~\ref{spectr} (from different emission bands)
implies that the medium, in the bulk, is significantly cooler than 1000\,K. 
This is the sort of temperature we have found for the disc-like structure in
Sect.~\ref{ire}.  However, this structure is expected to be mostly optically
thick.  It seems that possible sites for the emision line spectrum to
originate might be in regions between the stellar surface and the disc where
the contracting stellar envelope might have left some
matter circulating now near the equatorial plane.
\cite{banash} have found that the observed width of the KI 
emission lines if interpreted with Doppler broadening give a velocity of
260~km/s. The measured FWHM of the CaI line in our spectrum 
(see Table~\ref{spectr_t}) when corrected for the instrumental FWHM (5.6~\AA)
gives a similar value, i.e. 280~km/s. This value would correspond to a Keplerian
velocity at $\sim$10~$R_{\sun}$ for a 1~$M_{\sun}$ star. The required excitation 
could result from the stellar radiation or viscous
processes, or both. 

The galactic position of V4332~Sgr, $l = 13\fdg63$ $b = -9\fdg40$, might suggest 
that the object is related to
the Galactic bulge. Thus V4332~Sgr would be at a distance of $\sim$~8.5~kpc,
which, on the one hand, seems to be too large for the observed redenning, 
as discussed in Sect.~\ref{dist}.
On the other hand, however, it would be easier in this case to understand
the observed radial velocity of V4332~Sgr, which is, as discussed in Sect.~\ref{ire},
inconsistent with the rotation of the Galactic disc.
At a distance of 8.5~kpc all the radii and luminosities derived in this paper
would increase by a factor of 4.7 and 22, respectively. In particular,
the progenitor would be a G type subgiant of a mass of
$\sim$~2~$M_{\sun}$, as can be inferred from stellar evolutionary tracks
(e.g. Iben \cite{iben}), evolving from the main sequence to the red giant branch.
The star would have a $\sim$~0.2~$M_{\sun}$ helium core and an envelope
with a mean density of 0.02~g~cm$^{-3}$. Thus one could
refer to the scenario of Retter \& Marom (\cite{retmar}) while interpreting the
eruption of V4332~Sgr. Indeed $M_\mathrm{acc} \simeq 10^{-4} M_{\sun}$ is then
required to explain the energy emitted in the eruption so a massive planet or
a brown dwarf would be involved in the merger event. However,
the accreted object, being
significantly denser than the subgiant envelope, would be expected to penetrate
deeply in the envelope, as discussed e.g. in Livio \& Soker
(\cite{livsok}).  Thus most of the subgiant envelope would be
disturbed and the whole event, the contraction phase in particular,
would procede on a much longer time scale than observed. Indeed,
modelling analogous to that in Sect.~\ref{contrac} but done to match the observed
increase of the radius by a factor of 4.7, assuming
a central star mass of 2~$M_{\sun}$, requires $M_\mathrm{e} \simeq 3 \times 10^{-4}M_{\sun}$.
In the discussed case we would rather expect $M_\mathrm{e} \simeq 1 M_{\sun}$ 
to be involved. Note that Livio \&
Soker (\cite{livsok}) show that a merger event in the case of a giant and a
massive planet may last several thousands years.

There is no doubt that futher detailed observations of V4332~Sgr are
necessary. Photometric and spectroscopic measurements in the near 
and far infrared would be of particular value as the object is emitting
most of its energy in this wavelength range.  High quality spectroscopy 
in the optical would also be very important for understanding the origin of the 
unusual emission-line spectrum as well as for analysing the underlying stellar 
component.  In particular, high resolution spectroscopy in the atomic lines 
would allow precise determination of the radial velocity of the object, as well 
as investigation of the kinematic structure of the emitting region.
The object is probably evolving on a time scale of years so long term observational
monitoring should be undertaken.
We hope that the analysis of the existing observational data in this paper,
followed by our interpretation and conclusions (which in several points may be 
regarded as speculative) will stimulate deeper astrophysical interest in V4332~Sgr, 
V838~Mon and, possibly, other objects of similar nature.

\acknowledgements{The research reported in this paper has partly been supported
from a grant No. 2.P03D.002.25 financed by the Polish State Committee for
Scientific Research.}

\end{document}